\documentstyle[prl,aps,epsf]{revtex}
\begin{document}

\def \gam {\frac{ N_f N_cg^2_{\pi q\bar q}}{8\pi} }
\def \gamm {N_f N_cg^2_{\pi q\bar q}/(8\pi) }
\def \be {\begin{equation}}
\def \ba {\begin{eqnarray}}
\def \ee {\end{equation}}
\def \ea {\end{eqnarray}}
\def \gap {{\rm gap}}
\def \gapp {{\rm \overline{gap}}}
\def \gappp {{\rm \overline{\overline{gap}}}}
\def \im {{\rm Im}}
\def \re {{\rm Re}}
\def \Tr {{\rm Tr}}
\def \P {$0^{-+}$}
\def \S {$0^{++}$}
\def \uu {$u\bar u$}
\def \dd {$d\bar d$}
\def \ss {$s\bar s$}
\def \qq {$q\bar q$}
\def \qqq {$qqq$}
\def \si {$\sigma(500-600)$}
\def \lsm {L$\sigma $M}
\title{The $s \bar{s}$ and $K\bar K$ nature of $f_0(980)$ in $D_s$ decays}
\author{A. Deandrea$^{(a)}$, R. Gatto$^{(b)}$,
G.Nardulli$^{(c)}$,
A.D. Polosa$^{(d)}$ and N. A. T\"ornqvist$^{(d)}$}
\address{(a) Institut de Physique Nucl\'eaire, Universit\'e Lyon I,
43, bd du 11 Novembre 1918, F--69622 Villeurbanne Cedex, France
\\(b) D\'epartement de Physique
Th\'eorique, Universit\'e de Gen\`eve, 24 quai E.-Ansermet,
CH-1211 Gen\`eve 4, Switzerland \\
(c) Dipartimento di Fisica,
Universit\`a di Bari and INFN Bari, via Amendola 173, I-70126
Bari, Italy\\
(d) Physics Department, POB 9, FIN--00014, University of Helsinki,
Finland}
\date{December, 2000} \maketitle
\begin{abstract}
We examine the $D_s\to f_0(980)\pi$ amplitude through a
constituent quark-meson model, incorporating heavy quark and
chiral symmetries, finding a good agreement with the recent E791
data analysis of $D_s \to 3\pi$ via $f_0(980)$. The $f_0(980)$
resonance is considered at the moment of production as an $s
\bar{s}$ state, later evolving to a superposition of mainly $s \bar
s$ and $K \bar K$. The analysis is also extended to the more
frequent process $D_s\to \phi\pi$.
\\
\vskip 0.05cm \noindent Pacs numbers: 13.25.Ft, 12.39.Hg,
14.40.Cs\\ \noindent
LYCEN-2000-132\\ UGVA-DPT/2000-11-1090\\ HIP-2000-65/TH \vskip
0.90cm
\end{abstract}

{\it (i) Introduction}

The scalar mesons have remained a controversial enigma since a long time.
There is today no consensus as to the true nature of especially
the lightest scalars, the $\sigma$, $f_0(980)$ and $a_0(980)$. Are
these $q\bar q$ \cite{NAT}, $\pi\pi$ \cite{Oller}, $K\bar K$ bound
state by hyperfine interaction \cite{Weinstein} or multi-quark
\cite{Jaffe} states? Is the expected scalar gluonium state
\cite{Minkowski,Close} present among these mesons, either as the
dominant component, or through a small mixing?

These are fundamental questions of great importance in particle
physics. The mesons with vacuum quantum numbers are known to be
crucial for a full understanding of the symmetry breaking
mechanisms in QCD, and presumably also for confinement. The light
$\sigma$, near 500 MeV in mass and with a large width of the same
order of magnitude, has reappeared on the scene (it has been on
the short list of PDG since 1996 \cite{pdg96,pdg}), and a growing
number of analyses, using more sophisticated theoretical
techniques, now find this elusive meson in $\pi\pi\to\pi\pi$,
where its presence is almost hidden because  of the Adler zero (in
production experiments the $\sigma$  appears more clearly).
Recently, in June 2000, a conference, entirely devoted to the
$\sigma$, was held  in Kyoto \cite{kyoto}, and many new analyses
of old data were presented, showing a wide agreement on a $\sigma$
pole near $m-i\Gamma/2=500-i250$ MeV.

Experiments studying charm decay to light hadrons are opening up a
new unique experimental window for understanding light meson
spectroscopy, and especially the controversial scalar mesons. The
scalars are abundantly produced in these decays, and in the recent
E791 experiment \cite{E791I} the $\sigma$ contributes 46\% of the
$D\to 3\pi$, and $f_0(980)$ over 50\% of the $D_s\to 3\pi$ Dalitz
plot \cite{E791}.

$D_s$ decays are particularly interesting, since with the Cabibbo
favored $c\to s$ transition and the dominant spectator mechanism
one expects the final state to be dominated by $s\bar s$ states.
This mechanism is supported by the fact that, in $D_s$ decay, the
well known nearly-$s\bar s$ vector state, the $\phi(1020)$, is
abundantly produced,  and one consequently expects that also the
related scalar $s\bar s$ state  should appear strongly.
Since the $f_0(980)$ is produced copiously in $D_s$ decay, this
supports the picture of a large $s\bar s$ component in its
wave function.

Previously this predominant $s\bar s$ nature of the $f_0(980)$ has
been  supported by the radiative decay $\phi\to f_0(980)\gamma$
\cite{Achasov} and by unitarized quark models \cite{NAT}, although
being just below the $K\bar K$ threshold it should also have a
large component of virtual $K\bar K$ in its wave function.

In this paper we shall assume that both the $\phi (1020)$ and the
$f_0(980)$ are predominantly $s\bar s$ states, or at least that
when they are produced in $D_s$ decay, the production is via the
$s\bar s$ component. After production the $s\bar s $ core of
$f_0(980)$ induces a virtual cloud of $K\bar K$ (due to the fact
that $f_0(980)$ is just below the $K\bar K$ threshold and it
couples strongly to $K\bar K$) behaving as a large standing
$S-$wave surrounding the relatively small $q\bar q$ core. We
consider a model for heavy-light meson decays, the Constituent Quark
Meson Model (CQM) \cite{cqmtot}, so far successfully applied, and see if 
it predicts the decay rates for $D_s\to \phi\pi$ and
$f_0(980)\pi$ compatible with the recent data. The CQM has
been widely exploited for heavy-light meson decays, in which the
light quarks are $u$ or $d$.  Recently some of us
\cite{predicting} studied $D\to\sigma\pi\to 3\pi$ with the CQM,
finding good agreement with the E791 data for this process and
assuming the $\sigma$ is predominantly $(u\bar u+d\bar
d)/\sqrt{2}$. The role of the $\sigma$ in $B$ decays to 3 pions
was also investigated \cite{bsigma}.

In a recent paper by Anisovich {\it et al.}, the authors adopt the
hypothesis of an $f_0(980)$ having $s\bar{s}$ and $(u\bar u+d\bar
d)/\sqrt{2}$ components with a mixing angle of $-48^{o}$ deduced
from a phenomenological analysis of $\phi(1020)\to\gamma f_0(980),
\gamma \eta, \gamma \eta^\prime, \gamma \pi^0$ and $f_0(980)\to
\gamma \gamma$ radiative decays \cite{Anisovich}. The $K\bar K $
component of $f_0(980)$ is neglected in this approach. On the
other hand it has been shown by Markushin \cite{Markushin} how a
$K\bar K$ molecular picture of $f_0(980)$ can explain the
$f_0(980)\to \pi\pi$ decay with no need of $u\bar u, d\bar d$
components. As we shall see, our analysis favors a $f_0(980)$ 
produced as an $s \bar s$ state which
evolves generating a virtual cloud of $K \bar K$
eventually decaying in an OZI allowed way to $\pi\pi$, as shown in
Fig. 1. In fact we shall test the hypothesis that the $s\bar s$
component is substantial.

{\it (ii) $D_s \to f_0 \pi$}

The CQM model can be extended in the strange quark sector solving
the {\it gap equation} discussed in \cite{rass} with a non zero
current mass for the strange quark:
\begin{equation}
\Pi (m)=m - m_0 - 8m G I_1(m^2) = 0,
\end{equation}
where $G=5.25\;{\rm GeV}^{-2}$ and $m_0$ is the current mass of
the strange quark. The $I_1$ integral is calculated using the
proper time regularization:
\begin{equation}
I_1=\frac{iN_c}{16\pi^4} \int^{reg} \frac{d^4k}{(k^2 - m^2)}
={{N_c m^2}\over {16 \pi^2}} \Gamma\left(-1,{{{m^2}} \over
{{{\Lambda}^2}}},{{{m^2}}\over {{{\mu }^2}}}\right).
\end{equation}
The choice of the UV cutoff is dictated by the scale of chiral
symmetry breaking $\Lambda_\chi= 4 \pi f_\pi$ \cite{georgi} and we
adopt $\Lambda=1.25$ GeV. The IR cutoff $\mu$ and the constituent
mass $m$ must be fixed taking into account that CQM does not
incorporate confinement. This means that we have to enforce the
kinematical condition to produce  free constituent quarks $M \geq
m_Q + m$, where $M$ is the mass of the heavy meson and $m_Q$ is
the constituent mass of the heavy quark there contained.
Considering that the heavy meson momentum is $P^\mu=m_Q v^\mu +
k^\mu$, $v^\mu$ being the heavy quark 4-velocity and $k^\mu$ the
so called residual momentum due to the interactions of the heavy
quark with the light degrees of freedom at the scale of
$\Lambda_{\rm QCD}$, the above condition coincides with $v\cdot k
\geq m$ (since $P=Mv$, the 4-velocity of the meson is almost
entirely carried by the heavy quark), or equivalently, in the rest
frame of the meson, ${\rm inf}(k)=m$, meaning that the smallest
residual momenta that can run in the CQM loop amplitudes are of
the same size of the light constituent mass. The IR cutoff $\mu$
is therefore $\mu \simeq m$.

A reasonable constituent quark mass for the strange quark is
certainly $m=510$ MeV, considering the $\phi$ meson as a pure $s
\bar{s}$ state \cite{pdg}. Taking $\mu=0.51$ GeV as an infrared
cutoff, a value of $m_0=131$ MeV, see Fig. 2, is required by the
gap equation (consistently with the spread of values for the
current $s$ quark mass quoted into \cite{pdg}). Varying the
current strange mass in the range $60-170$ MeV reflects into a
small excursion of the constituent strange mass around the $500$
MeV value.

The free parameter of CQM is $\Delta_H$ defined by
$\Delta_H=M_H-m_Q$. The subscript $H$ refers to the $H-$multiplet
of Heavy-Quark-Effective-Theory (HQET) \cite{manowise}
$H=(0^-,1^-)$. In a similar way a $\Delta_S$ is associated to 
the $S$ multiplet, $S=(0^+,1^+)$.  
The latter is determined fixing $\Delta_H$
\cite{rass}. The related $\Delta_H,\Delta_S$ values in the strange
sector are shown in Table I. We consider the range of values
$\Delta_H=0.5,0.6,0.7$ GeV in all numerical computations. This
range is consistent with the condition $M \geq m_Q+m$.

As a first test we compute the decay constant $f_{D_s}$ given in
CQM by the following expression:
\begin{equation}
\label{eq:following}
f_{D_s}=\frac{\hat{F}}{\sqrt{m_{H}}}=\frac{2\sqrt{Z_H}
(I_1+(\Delta_H+m)I_3(\Delta_H))}{\sqrt{m_H}},
\end{equation}
where $m_H=m_{D_s}=1.968$ GeV and:
\begin{eqnarray}
I_3(\Delta) &=& - \frac{iN_c}{16\pi^4} \int^{\mathrm {reg}}
\frac{d^4l}{(l^2-m^2)(v\cdot l + \Delta + i\epsilon)}\nonumber \\
&=&{N_c \over {16\,{{\pi }^{{3/2}}}}}
\int_{1/{{\Lambda}^2}}^{1/{{\mu }^2}} {ds \over {s^{3/2}}} \; e^{-
s( {m^2} - {{\Delta }^2} ) }\; \left( 1 + {\mathrm {erf}}
(\Delta\sqrt{s}) \right).
\end{eqnarray}
The renormalization constant $Z_H$ is given by:
\begin{equation}
Z_H^{-1} = (\Delta_H+m) {\frac{\partial I_3(\Delta_H)}{\partial
\Delta_H}} +I_3(\Delta_H).
\end{equation}
Numerically we find:
\begin{equation}
f_{D_s}=297_{-22}^{+29}\;{\rm MeV},
\end{equation}
(the error is computed varying $\Delta_H$ in the range of values
quoted above) that is in good agreement with the value in the PDG
\cite{pdg}:
\begin{equation}
f_{D_s}=280 \pm 19\pm 28 \pm 34 \;{\rm MeV}.
\end{equation}

Let us now consider the $D_s\to f_0(980)$ semileptonic amplitude:
\begin{eqnarray}
\langle f_0(q_{f_0})|A^{\mu}_{(\bar{s}c)}(q)|D_s(p)\rangle &=&
\left[ (p+q_{f_0})^{\mu}+\frac{m_{f_0}^2-m_{D_s}^2}{q^2}q^\mu
\right]\; F_1(q^2)\nonumber \\ &-& \left[
\frac{m_{f_0}^2-m_{D_s}^2}{q^2}q^{\mu} \right] \; F_0(q^2),
\label{eq:effezero}
\end{eqnarray}
with $F_1(0)=F_0(0)$. This amplitude can be represented by the
diagram in Fig. 3. CQM allows to model the  $f_0$ vertex,
indicated with a black spot, through the diagrams in Figs. 4 and 5
respectively. The former gives what we call the {\it polar}
contribution to the form factors: considering the $0^-$
intermediate polar state one can compute $F_0$ while $1^+$ is
connected to $F_1$. The {\it direct} diagram, depicted in Fig. 5,
gives access to both the computation of $F_1$ and $F_0$. The
method and the computation technique have been fully explained in
\cite{predicting}. Actually, in a factorization scheme, the
amplitude describing the decay $D_s\to f_0(980)\pi$ is expressed
by the product of two matrix elements. One is the semileptonic
matrix element (\ref{eq:effezero}), the other is the well known:
\begin{equation}
\langle \pi|A^\mu_{(\bar{u}d)}(q)|{\rm VAC}\rangle=if_\pi q^\mu.
\label{eq:pcac}
\end{equation}
The $q^\mu$ in this matrix element selects $F_0$ as the relevant
form factor for the computation of the amplitude. The polar and
direct contributions to the form factors must be added in order to
determine $F_0(q^2=m_\pi^2\simeq 0)$. In the case at hand we use a
coupling of $f_0(980)$ to the light quarks that, for $SU_3-$flavor
symmetry, is $g_{f_0 ss}=\sqrt{2} g_{\sigma q q}= 2.49\sqrt{2}$
\cite{ebertbos}. This of course comes from the hypothesis of an
$f_0(980)$ having a $\bar{s} s$ structure. If on the other hand
one would adopt the picture given in \cite{Anisovich}, $g_{f_0
ss}$ should be reduced by a factor of sin$(\phi)$, with
$\phi=-48^o$. The numerical value for $F_0$ is obtained through
the CQM expressions containing the $I_i(\Delta_H)$ and
$I_i(\Delta_S)$ integrals discussed in \cite{predicting}. The
result is the following:
\begin{equation}
F_0(q^2=0)=F_0^{({\rm pol})}(0)+ F_0^{({\rm
dir})}(0)=0.64_{-0.03}^{+0.05}.
\end{equation}
The expression for the decay amplitude is:
\begin{equation}
g_{D_s f_0 \pi}=\langle f_0 \pi^+|H_{\rm
eff}|D^+_s\rangle=\frac{G_F}{\sqrt{2}}V^*_{cs}V_{ud}a_1
F_0(0)(m_{D_s}^2-m_{f_0}^2)f_\pi,
\end{equation}
where $H_{\rm eff}$ is the effective Hamiltonian of Bauer, Stech
and Wirbel \cite{BSW}, with $a_1=1.10\pm 0.05$ fitted for $D$
decays \cite{nonlep}. Its numerical value is:
\begin{equation}
g_{D_s f_0 \pi}=(2.08_{-0.12}^{+0.16})\times 10^{-6}\;{\rm GeV},
\end{equation}
and therefore  the predicted width is:
\begin{equation}
\label{eq:lambda} \Gamma (D_s^+ \to f_0(980) \pi^+)=\frac{g_{D_s
f_0 \pi}^2}{16\pi m_{D_s}^3}\sqrt{\lambda
(m_{D_s}^2,m_{f_0}^2,m_\pi^2)}=(3.27_{-0.35}^{+0.52})\times
10^{-14}\;{\rm GeV},
\end{equation}
($\lambda$ is the triangular function), to be compared with the
PDG \cite{pdg} one:
\begin{equation}
\Gamma (D_s^+ \to f_0(980) \pi^+)=(2.39\pm 1.06) \times
10^{-14}\;{\rm GeV}. \label{eq:speri}
\end{equation}
Considering the mixing of $s \bar s$ with the $(u \bar u + d \bar
d)/\sqrt{2}$ component, i.e., using the reduced coupling $g_{f_0
ss} \sqrt{2}$sin$(\phi)$ discussed above, one would obtain a
different prediction for the width:
\begin{equation}
\Gamma(D_s\to f_0(980)\pi)=(1.86^{+0.28}_{-0.25})\times 10^{-14}.
\label{eq:lambda1}
\end{equation}
Even if (\ref{eq:lambda}) and (\ref{eq:lambda1}) seem both to
agree with the experimental value (\ref{eq:speri}), for reasons
that will be explained soon, the CQM model definitely favors the
$s\bar s$ scenario.

In order to make a comparison with E791 results \cite{E791},
we compute the branching ratio ${\cal B}(D_s^+ \to f_0(980)
\pi^+\to 3\pi)$ estimating the coupling $g_{f_0 \pi^+ \pi^-}$ for
$f_0\to \pi^+\pi^-$ through the following formula:
\begin{equation}
\frac{g_{f_0 \pi^+ \pi^-}^2}{4\pi}=\frac{2 m_{f_0}^2 C
\Gamma_0}{\sqrt{m_{f_0}^2/4-m_\pi^2}},
\label{ci}
\end{equation}
where $C=2/3\times 0.68 \simeq 4/9$ is the fraction of the total
$f_0(980)$ width related to the process $f_0\to\pi^+\pi^-$ (the
$K\bar{K}$ decay mode is also considered, $2/3$ being the isospin
factor and $0.68=\Gamma(f_0\to \pi\pi)/(\Gamma(f_0\to
\pi\pi)+\Gamma(f_0\to K\bar K))$ \cite{pdg}) and $\Gamma_0=44$ MeV
is the central value of the width of $f_0(980)$ found by E791
\cite{E791}. Using the expression:
\begin{equation}
\Gamma(D_s^+\to f_0(980)\pi^+\to
3\pi)=\frac{1}{2}\int_{4m_\pi^2}^{(m_{D_s}-m_\pi)^2} ds \;
\Gamma_{D_s\to
f_0\pi}(s)\times\frac{1}{\pi}\frac{\sqrt{s}\Gamma_{f_0\to
\pi^+\pi^-}(s)}{(s-m_{f_0}^2)^2+m_{f_0}^2 \Gamma_{f_0}^2(s)},
\label{eq:mado}
\end{equation}
where the $\Gamma$'s are computed assuming a pure $s\bar s$
component in production as in (\ref{eq:lambda}) but substituting
$m_{f_0}^2\to s$, and the co-moving width is:
\begin{equation}
\Gamma_{f_0}(s)=\Gamma_0\times
 \frac{m_{f_0}}{\sqrt{s}}\frac{\sqrt{s/4-m_\pi^2}}{\sqrt{m_{f_0}^2/4-m_\pi^2}}.
\end{equation}
Assuming the E791 value $m_{f_0}=975$ MeV \cite{E791} and taking
the central value of the PDG branching ratio ${\cal B}(D_s\to
3\pi)=1.0\%$ (disregarding  the large given uncertainty $\pm 0.4$)
one finds that the fraction of $D_s$ decaying into $3\pi$ via
$f_0(980)$ is given by:
\begin{equation}
{\cal B}(D_s^+ \to f_0(980) \pi^+\to 3\pi)=50_{-6}^{+8}\;\%.
\end{equation}
This turns out to be in good agreement with the E791 results
\cite{E791}:
\begin{equation}
{\cal B}(D_s^+ \to f_0(980) \pi^+\to 3\pi)=(56.5\pm 4.3 \pm
4.7)\;\%,
\end{equation}
as far as the central value of the experimental branching ratio
${\cal B}(D_s\to 3\pi)$ is concerned. Unfortunately the lack of
precision in this measurement weakens the comparison between the
CQM calculation and the E791 results. 

{{\it (iii)} $D_s\to \phi \pi$}

In order to overcome the
ambiguity between what found in the pure $s \bar s$ hypothesis and
in the case of an $f_0(980)$ pictured as a mixture of $s\bar s$
and $(u\bar u + d\bar d)/\sqrt{2}$ (both results (\ref{eq:lambda})
and (\ref{eq:lambda1}) seem to be consistent with the experimental
value (\ref{eq:speri})) we consider the CQM analysis of the
$D_s\to\phi\pi$ decay in order to compare the widths $\Gamma
(D_s\to \phi\pi)$ and $\Gamma (D_s\to f_0 \pi)$. We believe that
the ratio $\Gamma (D_s\to f_0 \pi)/\Gamma (D_s\to \phi\pi)$ is the
most reliable theoretical output of our approach to be compared
with experimental data since it is less dependent on CQM
parameters and experimental normalizations. We will find that also
the $\Gamma (D_s\to \phi\pi)$ turns out to be larger with respect
to the experimental one, analogously to what found in
(\ref{eq:lambda}) with respect to (\ref{eq:speri}). Interestingly
the ratio of the two widths is in very good agreement with the
experimental ratio. This agreement is instead destroyed if a large
mixing with $u \bar u$ and $d \bar d$ components is taken into
account.

Considering the $\phi(1020)$ as an $\bar{s}s$ state, we can again
make use of factorization hypothesis \cite{BSW} to compute the
$D_s\to \phi \pi$ amplitude. The semileptonic form factors are
defined by:
\begin{eqnarray}
\langle \phi(\epsilon,
p^\prime)|A^\mu_{(\bar{s}c)}(q)|D_s^+(p)\rangle &=&
(m_{D_s}+m_\phi)A_1(q^2)\epsilon^{*\mu}\nonumber\\
-\frac{(\epsilon^{*}\cdot q)}{(m_{D_s}+m_\phi)}(p+p^\prime)^\mu
A_2(q^2)&-& (\epsilon^{*}\cdot q)\frac{2m_\phi}{q^2}q^\mu
(A_3(q^2)-A_0(q^2)).
\end{eqnarray}
To avoid the singularity in $q^2=0$, the condition:
\begin{equation}
\label{eq:prima}
A_0(0)=A_3(0),
\end{equation}
must hold. Moreover:
\begin{equation}
\label{eq:seconda}
A_3(q^2)=\frac{m_{D_s}+m_\phi}{2 m_\phi}
A_1(q^2)- \frac{m_{D_s}-m_\phi}{2 m_\phi}A_2(q^2).
\end{equation}
Observing that $q=(p-p^\prime)$ and using eq. (\ref{eq:pcac}) one
obtains:
\begin{equation}
g_{D_s \phi \pi}=\langle \phi \pi^+|H_{\rm
eff}|D^+_s\rangle=\frac{G_F}{\sqrt{2}}V^*_{cs}V_{ud}a_1
A_0(m_\pi^2) 2m_\phi (\epsilon^* \cdot q)f_\pi.
\end{equation}
In order to compute the width, a sum over final $\phi$
polarizations is performed. The form factor $A_0(0)$ can be
computed with CQM following the approach outlined in
\cite{nostrorho}. There is a {\it direct} and a {\it polar}
contribution to the form factors, as in the diagrams in Figs. 4,5.
The $\phi$ particle attached to the loop is introduced, via
Vector-Meson-Dominance (VMD), through an interpolating current:
\begin{equation}
J_\mu=\frac{m_\phi^2}{c_\phi f_\phi}\gamma_\mu,
\end{equation}
where the leptonic decay constant $f_\phi$ is defined by:
\begin{equation}
\langle {\rm VAC}| V_\mu|\phi(\epsilon)\rangle = f_\phi
\epsilon_\mu,
\end{equation}
with $V_\mu=-\bar s \gamma_\mu s$, and:
\begin{equation}
c_\phi=\frac{1}{\sqrt{6}}{\rm cos}(\theta_{\phi\omega}),
\end{equation}
$\theta_{\phi\omega}=39.4^o$ being the $\phi-\omega$ mixing angle,
see e.g. \cite{ballafre}. This current appears in the CQM loop
integral calculation in correspondence of the vertex (light
quark)-($\phi$)-(light quark). The lepton decay constant $f_\phi$
can be extracted from the value of the width:
\begin{equation}
\Gamma(\phi\to e^+e^-)=\frac{4\pi\alpha^2}{3}\left(\frac{f_\phi
c_\phi}{m_\phi^2}\right)^2 m_\phi,
\end{equation}
(as is checked with a straightforward calculation), given in the
PDG \cite{pdg}. We will use the value $f_\phi=249\;{\rm MeV}^2$.

Following \cite{nostrorho}, the direct contribution to $A_0(0)$ is
given by:

\begin{eqnarray}
A^{{\rm dir}}_0 (q^{2}=0) &=& -\frac{m_\phi }{f_\phi c_\phi}
\sqrt{Z_H m_{D_s}} \left[\Omega_1 \left(  m_\phi {\bar{\omega}}
-\frac{r_1}{m_{D_s}}\right) +  m_\phi \Omega_2 + \nonumber
\right.\\ &&\left. 2\Omega_3 + \Omega_4
+ \Omega_5 + 2\Omega_6 \left(  \bar{\omega}- \frac {r_1}{m_{D_s}
m_\phi} \right)-\nonumber\right.\\ &&\left.  Z \left(m^2  - m
\frac{r_1}{m_{D_s}}  + m m_\phi {\bar{\omega}}\right) \right],
\end{eqnarray}
where:
\begin{eqnarray}
\bar{\omega}&=&\frac{m_{D_s}^2+m_\phi^2}{2
m_{D_s}m_\phi}\nonumber\\ r_1 &=& \frac{m_{D_s}^2-m_\phi^2}{2},
\nonumber
\end{eqnarray}
and the integrals $Z, \Omega_j $, which are functions of
$\Delta_1$, $\Delta_2$, $x$ and $\bar{\omega}$, are tabulated in
\cite{nostrorho} and computed using $x=m_\phi$, the $\bar{\omega}$
given above and the $\Delta_H$ values from Table I. $\Delta_2$ are
here substituted by $\Delta_H-m_\phi \bar{\omega}$. The polar form
factors needed to compute $A_0^{\rm pol}(q^2=0)$ are, according to
eqs. (\ref{eq:prima}) and (\ref{eq:seconda}) , $A_1^{\rm pol}(0)$
and $A_2^{\rm pol}(0)$, given respectively by \cite{nostrorho}:

\begin{eqnarray}
A^{\rm pol}_1 (0) &=& \frac{\sqrt{2 m_{D_s}}g_V {\hat
F}^+}{m_{D_s(1^+)} (m_{D_s}+m_{\phi})} (\zeta-2\mu {\bar \omega}
m_{\phi})\\ A^{\rm pol}_2 (0) &=& -\sqrt{2} g_V \mu {\hat F}^+
\frac{\sqrt{m_{D_s}} (m_{D_s}+m_\phi)} {m_{D_s(1^+)}^2},
\end{eqnarray}
where now $\bar{\omega}=\frac{m_{D_s}}{2 m_\phi}$,
$g_V=\frac{m_\phi}{f_\pi}$ and $\hat{F}^+$ is defined by:
\begin{equation}
\langle {\rm VAC}|A_\mu|D_s(1^+)\rangle=i\sqrt{m_{D_s(1^+)}} v^\mu
{\hat F}^+,
\end{equation}
and computed in analogy with (\ref{eq:following}), see
\cite{rass}. The $\zeta$ and $\mu$ strong coupling constants are
described in \cite{nostrorho}:
\begin{eqnarray}
\mu &=& \frac{m^2_\phi}{\sqrt{2} g_V f_{\phi} c_\phi} \sqrt{Z_H
Z_S}\left( -\Omega_1- 2 \frac{\Omega_6}{m_\phi}+ m Z\right) \\
\zeta &=& \frac{\sqrt{2} m^2_\phi}{g_V f_{\phi}c_\phi} \sqrt{Z_H
Z_S} \left(m_\phi \Omega_2 +2 \Omega_3 +\Omega_4 +\Omega_5 -m^2 Z
\right),
\end{eqnarray}
where the $\Omega_j$ integrals involved are functions of
$\Delta_1=\Delta_H$, $\Delta_2=\Delta_S$, $x=m_\phi$ and
$\omega=(\Delta_1-\Delta_2)/{m_\phi}$. The renormalization
constant $Z_S$ is defined by:
\begin{equation}
Z_S^{-1} = (\Delta_S-m) {\frac{\partial I_3(\Delta_S)}{\partial
\Delta_S}} +I_3(\Delta_S).
\end{equation}
Of course everywhere $m=510$ MeV is the constituent mass of the
strange quark. $\mu$ and $\zeta$ are affected
by a considerable uncertainty due to varying $\Delta_H$ in the range of
Table I; this is reflected in a large uncertainty in the polar
form factor $A_0^{\rm pol}(0)=-0.16^{+0.27}_{-0.2}$. The direct
form factor is instead more stable against $\Delta_H$ variations
being $A_0^{\rm dir}(0)= 1.26^{+0.19}_{-0.15}$. Considering that:
\begin{equation}
A_0(0)=A_0^{\rm pol}(0)+A^{\rm dir}_0(0),
\end{equation}
we can readily compute the CQM ratio of widths in the case of only $s\bar s$ 
in the production: 
\begin{equation}
R=\frac{\Gamma(D_s^+\to f_0\pi^+)}{\Gamma(D_s^+\to \phi
\pi^+)}=0.4\pm 0.21,
\end{equation}
to be compared with the PDG one \cite{pdg}:
\begin{equation}
R=\frac{\Gamma(D_s^+\to f_0\pi^+)}{\Gamma(D_s^+\to \phi
\pi^+)}=0.49 \pm 0.20. \label{eq:mado2}
\end{equation}

{\it (iv) Conclusions}

If one adopts the hypothesis of Anisovich {\it et al.}, the
ratio $R$ would be reduced to $R=0.22\pm 0.12$. E791  also
measures the ratio $\Gamma(D_s\to 3\pi)/\Gamma(D_s \to f_0(980)
\pi)=0.245$ \cite{E791} with a very small uncertainty. If one
considers (\ref{eq:mado}) in the limit of narrow width for
$f_0(980)$, one obtains $\Gamma(D_s\to f_0(980)\pi\to 3\pi)= C
\; \Gamma (D_s\to f_0(980)\pi)$ (see the discussion after 
eq.\ref{ci}). On the other hand E791 finds that
$\Gamma(D_s\to f_0(980)\pi\to 3\pi)=56.5 \% \Gamma(D_s\to 3\pi)$
or, in other words, they measure $R=0.62$ with a very small error.
This indeed agrees with the known PDG result (\ref{eq:mado2}).

The computed widths $D_s\to f_0\pi$ and $D_s\to \phi \pi$ are both
larger with respect to the corresponding experimental values,
nevertheless their ratio is only $20\%$ smaller than the
experimentally estimated ratio. Our results
favor the scenario of an $f_0(980)$ made of an $s\bar{s}$ core
surrounded by a standing $S$-wave of virtual $K \bar
K$. A large $u\bar u$, $d \bar d$ component in $f_0(980)$ seems
also excluded by the fact that in $D\to 3\pi$ decays the
$f_0(980)$ is weakly produced \cite{E791I}.

Therefore, in conclusion, our work supports a description of $f_0(980)$ as
$s\bar{s}$, with virtual $K \bar K$ cloud. Any substantial mixture
of $u\bar u$, $d \bar d$ seems excluded. Light-quark phenomenology
has been fighting since a long time to understand the scalar mesons.
It now appears that heavy meson decays may be able to clarify
this difficult problem: {\it nemo propheta in patria}.

\twocolumn
\acknowledgements
ADP and NAT acknowledge support  from EU-TMR
program, contract CT98-0169.

\begin{table}
\hfil \vbox{\offinterlineskip \halign{&#&
\strut\quad#\hfil\quad\cr \hline \hline &$\Delta_H$ && $\Delta_S$&
\cr \hline &$0.5$&& $0.86$ &\cr &$0.6$&& $0.91$ &\cr &$0.7$&&
$0.97$ &\cr \hline \hline }} \caption{$\Delta$ values in (in GeV)}
\label{t:senza1}
\end{table}

\begin{figure}
\epsfxsize=7cm \centerline{\epsffile{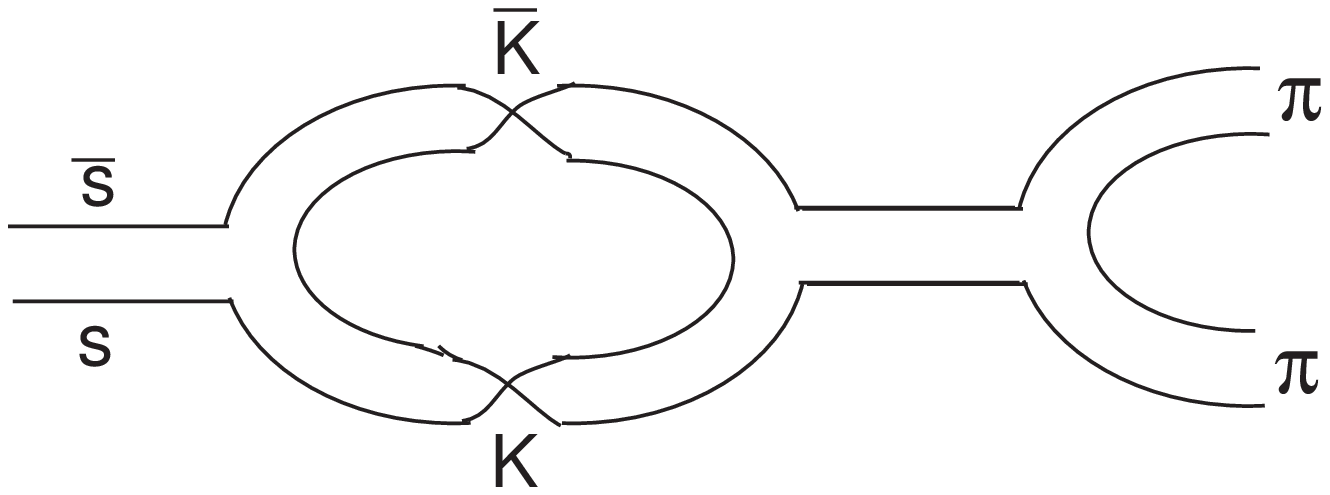}} \noindent {\bf
Fig. 1} - {After the production at short distances in $D_s$ decay
via its $s\bar s$ component (see Figs. 3-5), the $f_0(980)$, being
just below the $K\bar K$ threshold, evolves in time generating a
substantial $K \bar K$ component (with larger spatial dimension
than $s \bar s$) that can decay, in OZI allowed way, to $2\pi$.}
\end{figure}

\begin{figure}
\epsfxsize=7cm \centerline{\epsffile{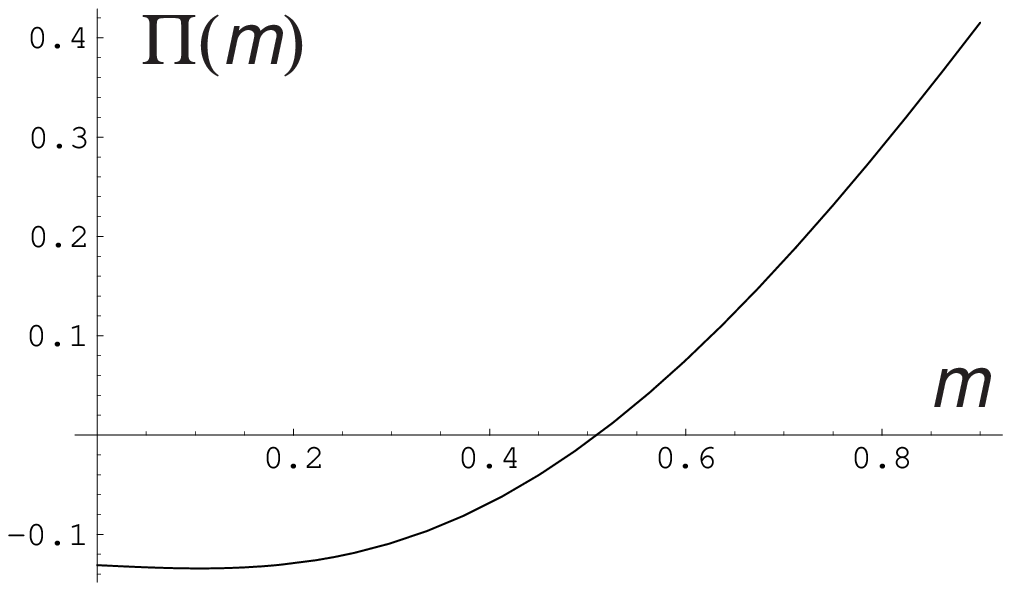}} \noindent {\bf
Fig. 2} - {Gap equation zero for a fixed value of $m_0$. The
masses are expressed in GeV. Here $m_0=131$ MeV is the current
mass of the {\it strange} quark.}
\end{figure}

\begin{figure}
\epsfxsize=7cm \centerline{\epsffile{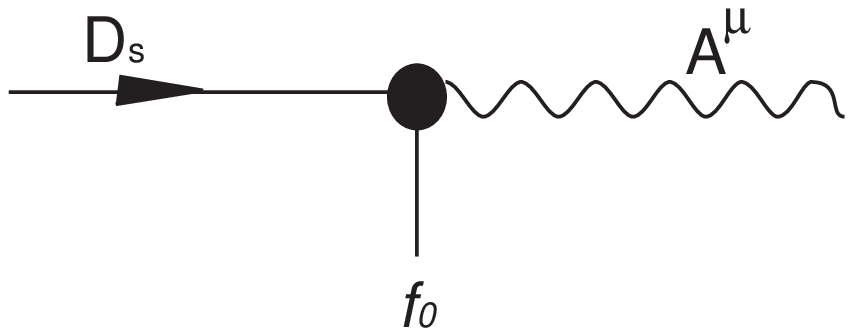}} \noindent {\bf
Fig. 3} - {The semileptonic amplitude. The vertex with the weak
current and the $f_0$ can be modeled with CQM as is described in
Figs. 3 and 4. The same diagrams with $\phi$ in place of $f_0$ are
also considered. The $\phi$ resonance is introduced, via VMD,
through an interpolating current $J_\mu$.}
\end{figure}

\begin{figure}
\epsfxsize=7cm \centerline{\epsffile{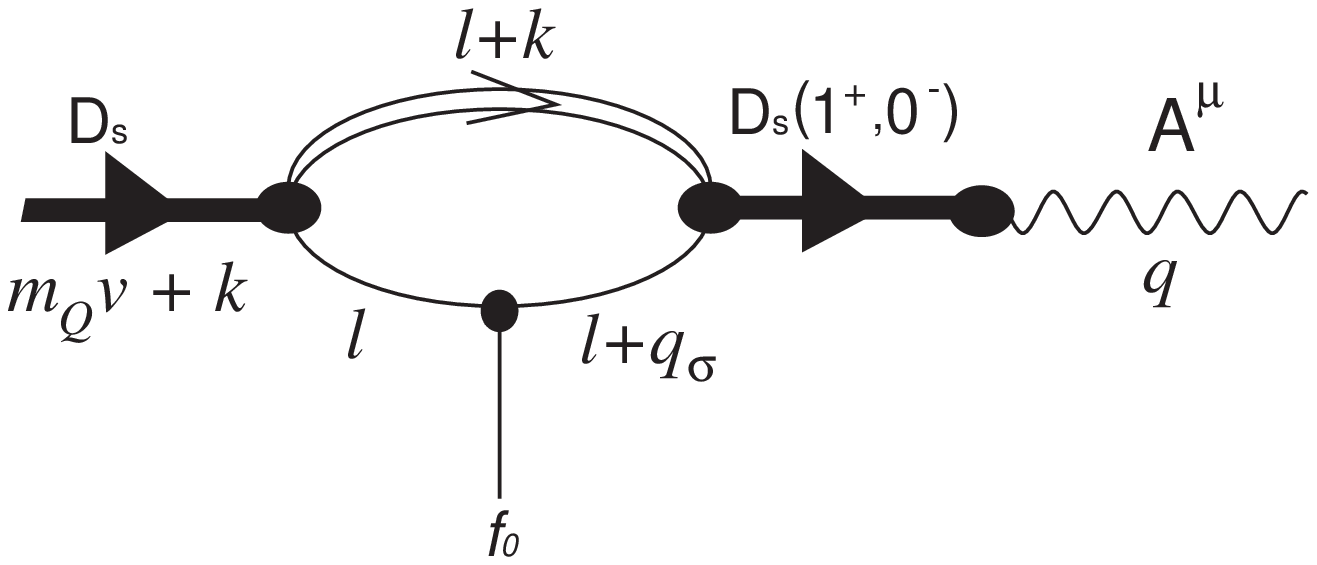}} \noindent {\bf
Fig. 4} - {The {\it polar} diagram. The polar contribution to the
form factor is reliable when computed near the pole mass. The
uncertainty in the extrapolation $q^2\to 0$ reflects in the
violation of the condition $F_0^{\rm pol}(0)=F_1^{\rm pol}(0)$.
This kind of uncertainty is taken into account in our
calculation.}
\end{figure}

\begin{figure}
\epsfxsize=7cm \centerline{\epsffile{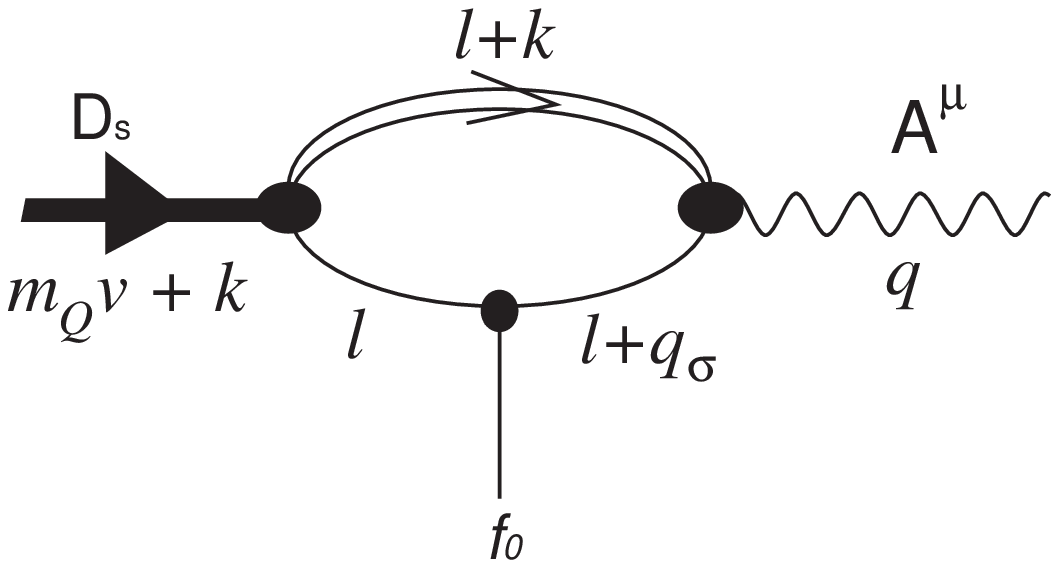}} \noindent {\bf
Fig. 5} - {The {\it direct} diagram. The condition $F_0^{\rm
dir}(0)=F_1^{\rm dir}(0)$, avoiding the spurious singularity in
$q^2=0$ in (\ref{eq:effezero}), is automatically satisfied.}
\end{figure}

\end{document}